\documentclass{moriond}

\def\be{\begin{equation}}
\def\ee{\end{equation}}
\def\bea{\begin{eqnarray}}
\def\eea{\end{eqnarray}}

\newcommand{\Photo}{\includegraphics[height=35mm]{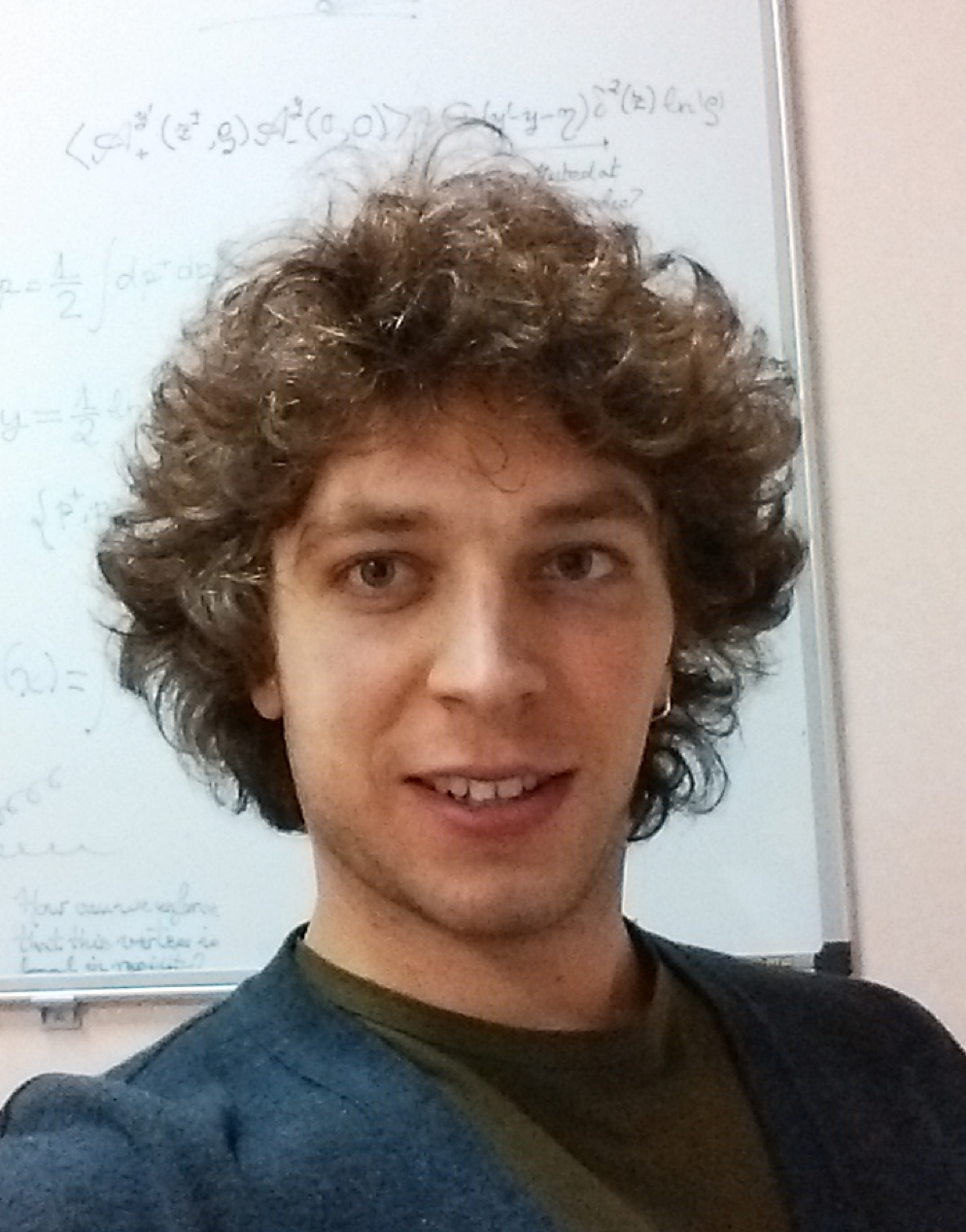}}

\begin{document}
\vspace*{4cm}
\title{The neutron EDM vs up and charm flavour violation}

\author{Filippo Sala}

\address{\normalsize{\it Institut de Physique Th\'eorique, CNRS and CEA/Saclay, F-91191 Gif-sur-Yvette, France}}

\maketitle\abstracts{
We derive a strong bound on the chromo-electric dipole moment of the charm quark, and we quantify its impact on models that allow for a sizeable flavour violation in the up quark sector%, like flavour alignment and Generic $U(2)^3$
.
In particular we show how the constraints coming from the charm and up CEDMs limit the size of new physics contributions to direct flavour violation in $D$ decays. We also specialize our analysis to the cases of split-families Supersymmetry and composite Higgs models. The results we expose motivate an increase in experimental sensitivity to fundamental hadronic dipoles, and a further exploration of the SM contribution to both flavour violating $D$ decays and nuclear electric dipole moments.}

\section{Introduction and motivation}
\label{sec:introduction}
The Cabibbo Kobayashi Maskawa (CKM) picture of flavour and CP violation in the quark sector has successfully predicted all the related experimental results to date. The high level of precision to which it has been tested can be quantified by parameterising new physics (NP) via the effective Lagrangian
\be
\label{eq:eff_L}
\mathcal{L}_{\rm NP}=\sum_i \frac{c_i}{\Lambda_i} \mathcal{O}_i\, ,
\ee
where $\mathcal{O}_i$ are flavour violating (FV) operators of dimension six, that describe the effect of new degrees of freedom appearing roughly at the scale $\Lambda_i$. If one assumes the coefficients $c_i$ to be of order one, as expected for a generic NP model, then the bounds on such scales $\Lambda_i$ can reach values of $10^{4\div 5}$ TeV, depending on the operator under consideration.
Any theory that requires the existence of new states below these values has to face this issue, which is often referred to as the ``NP flavour problem". In particular, such high scales are unacceptable for models that aim at addressing the hierarchy problem in a \textit{natural} way, since by construction their new scale is as close as possible to the electroweak (EW) one%\footnote{but notice that these scales also have some relevance for theories where the hierarchy problem is addressed in other ways (e.g. anthropically)%, like some versions of Split-Supersymmetry
%.}
.
Actually these models strictly require only the NP scale associated with the third generation to be close to the Fermi one, while the scale associated with the first two generations could be well above it. This separation is indeed a welcome property for natural models in light of collider constraints.

A possible way to address the NP flavour problem is to give the theory a specific feature, like a flavour symmetry or some dynamical property, that makes all the desired coefficients $c_i$ small enough.
%A widely studied case is to impose large flavour groups ($U(3)^3$, $U(2)^3$) broken minimally along specific directions. In this way flavour violation is not only under control, but also "CKM-like", i.e. only the Standard Model (SM) operators are important and for example FV effects in the up quark sector are well below experimental sensitivities.
In doing so, it can be useful to keep in mind that the higher energy scales in eq. (\ref{eq:eff_L}) are probed by operators involving down quarks, strongly constrained by measurement of decay properties of $K$ and $B$ mesons. On the contrary, constraints from the up-quark sector probe lower scales. Despite this, it is not difficult to build models where flavour violation in the up sector is in the ballpark of current experimental sensitivities, like for example flavour alignment \cite{Nir:1993mx}, composite Higgs models with an anarchic flavour structure, and Generic $U(2)^3$ \cite{Barbieri:2012bh}.

In light of the previous discussion, it is interesting to look for indirect experimental signatures of a class of models defined by %the following properties
i) NP scale of the third generation below the one of the first two, and ii) largish FV effects in the up quark sector.
%\begin{itemize}
%\item[i)] NP scale of the third generation below the one of the first two, and
%\item[ii)] largish FV effects in the up quark sector.
%\end{itemize}
Within this framework, operators of particular interest are the dipoles
\be
\label{eq:upDipole}
\mathcal{L}_{\rm NP} \supset c_{ij} \frac{m_t}{\Lambda^2} (\bar{u}_{Li} \sigma_{\mu\nu} T^a u_{Rj}) g_s G_{\mu\nu}^a\, ,
\ee
where $i,j = u,c$. The strongest experimental probes of the associated energy scale are, currently, direct CP asymmetries in $D$ meson decays \cite{Isidori:2011qw}, $A_{\rm CP}$, and the neutron EDM $d_n$. The former are sensitive to $c_{uc}$ and $c_{cu}$, and the latter to $c_{uu}$, cause $c_{uu}$ is directly proportional to the chromo-EDM of the up quark. Property i) of our class of models implies the relation
\be
\label{eq:coeffs}
c_{uu} c_{cc} = c_{uc} c_{cu}\,,
\ee
so that one can have a largish contribution to A$_{\rm CP}$ and respect at the same time the $d_n$ bound, by taking a large enough $c_{cc}$.%This property has been widely used in previous studies (e.g. \cite{Giudice:2012qq}).
 
In this contribution, which is based on \cite{Sala:2013osa}, we will show that the neutron EDM is actually sensitive to $c_{cc}$. The constraint we will derive, despite being a first bound on the charm chromo-EDM, has important phenomenological consequences for all models satisfying eq.(\ref{eq:coeffs}): the bound from $d_n$ challenges the possibility to observe an asymmetry A$_{\rm CP}$. This conclusion will be remarkably strengthened by the expected future progresses, as we will now explain.
%To understand this statement more quantitatively, it can be useful to recall that t
The current upper bound on the neutron EDM reads \cite{Baker:2006ts} $d_n < 2.9 \times 10^{-26} e$ cm (90\% CL), and the foreseen experimental sensitivity is at the level of $10^{-28} e $ cm, and is aimed at by many experimental collaborations (see e.g. section 7 of \cite{Hewett:2012ns} and references therein)%\cite{Bodek:2008gr,Altarev:2009zz,Baker:2010zza}
. Such a value is a genuine probe of new physics, since the SM contribution is estimated to be at the level of $10^{-31} e $ cm \cite{Mannel:2012qk}.
The stronger constraint on CPV decays of the $D$ meson comes from the LHCb measurement \cite{Aaij:2013bra} of the difference $\Delta$A$_{\rm CP} = $A$_{\rm CP}(D \to K^+ K^-) - $A$_{\rm CP}(D \to \pi^+\pi^-)$, whose world average \cite{Amhis:2012bh} is $(-3.29 \pm1.21)\times10^{-3}$. The SM prediction for such decays is at the level of a few$\times 10^{-3}$, and is plagued by long distance uncertainties that are still a source of discussion in the community \cite{Pirtskhalava:2011va,Cheng:2012wr,Brod:2012ud,Isidori:2012yx}.

In the rest of this contribution we will derive the bound in section \ref{sec:bound}, and analyse quantitatively its phenomenological implications in section \ref{sec:pheno}. A critical summary and an outlook will be provided in section \ref{sec:outlook}.

\section{Bound on the chromo-EDM of the charm quark}
\label{sec:bound}
The expression of the neutron EDM in terms of fundamental quantities, if a Peccei-Quinn symmetry is assumed to get rid of the $\theta_{\rm QCD}$ term, reads \cite{Pospelov:2000bw}
\be
\label{eq:dn}
d_n = (1 \pm 0.5) \big[1.4 (d_d - 0.25 d_u) + 1.1 e (\tilde{d}_d + 0.5 \tilde{d}_u) \big] \pm (22 \pm 10)e\,{\rm MeV}\,w \,,
\ee
where $d_{u,d}$ and $\tilde{d}_{u,d}$ are the EDM and CEDM of the up and down quarks, and $w$ is the coefficient of the three-gluon Weinberg operator. They are defined via the effective Lagrangian \footnote{With the definition of the effective operator in (\ref{eq:upDipole}), one has $\tilde d_q = 2 \, (m_t/\Lambda^2) {\rm Im}(c_{qq})$.}
\be
\label{dipoles_def}
\mathcal{L}_{\rm eff} = d_q \,\frac{1}{2} (\bar{q} \sigma_{\mu\nu} i \gamma_5 q) F^{\mu\nu} + \tilde{d}_q \,\frac{1}{2} (\bar{q} \sigma_{\mu\nu} T^a i \gamma_5 q) g_s G_a^{\mu\nu} + w \,\frac{1}{6} f^{abc} \epsilon^{\mu \nu \lambda \rho} G^a_{\mu\sigma} G^{b\,\sigma}_\nu G^c_{\lambda\rho}\,,
\ee
where $q = u,d,s,c,b,t$ and $\epsilon^{0123}=1$. Every time we will employ the $d_n$ bound to constrain the size of new physics, we will conservatively use the values 0.5 and 12 MeV for the coefficients $1 \pm 0.5$ and $22 \pm 10$ MeV in (\ref{eq:dn}).

%\begin{equation}
%d_{q} = 2 e \, \frac{m_t}{\Lambda^2}\, {\rm Im}(c_{qq}),\qquad \tilde d_q = 2 \, \frac{m_t}{\Lambda^2}\, {\rm Im}(c_{qq})\,,
%\end{equation}
% Ignoring this assumption would introduce of course a strong dependence on this parameter, and also change the combination of $\tilde{d}_{u,d}$ on which $d_n$ depends, however keeping the order of magnitude of their impact.
The quarks EDMs and CEDMs do not mix in the Weinberg operator via renormalisation group evolution, however they nonetheless give a contribution to $w$. In fact, when in the running from high to low energies a heavy quark is integrated out, its CEDM gives a threshold correction to the Weinberg operator that reads \cite{Chang:1991ry,Boyd:1990bx,Dine:1990pf}
\be
\label{Weinberg_threshold}
w = \frac{g_s^3}{32 \pi^2} \frac{\tilde{d}_q}{m_q}\,,
\ee
where all the parameters are evaluated at the mass of the quark $q$. Expression (\ref{Weinberg_threshold}) is the one-loop result, the uncertainty coming from higher loops can be estimated to be at the level of $8 \alpha_s (m_q)/4\pi$($\simeq 25 \%$ for $q=c$), where 8 is a colour factor.
The subsequent running from $m_q$ makes also the lighter quarks dipole moments sensitive to $\tilde{d}_q$. For the charm quark, the impact of $\tilde{d}_c$ in $d_n$ is mainly driven by its contribution to $w$, which dominates over the light quark dipole moments by roughly two order of magnitudes. % The same is true e.g. for the top CEDM, where however the light quarks DMs contribute a the level of 10\%.
Using the threshold contribution (\ref{Weinberg_threshold}) and the one-loop running from \cite{Braaten:1990gq,Degrassi:2005zd}, one can write the expression
\be
\label{eq:dn_dc}
d_n = d_n(d) \pm d_n(w), \quad \, d_n(d) = (1\pm 0.5) (4.9 \times 10^{-6} e\, \tilde{d}_c), \quad \, d_n(w) = (1\pm 0.45) (5.1 \times 10^{-4} e\, \tilde{d}_c),
\ee
where $\tilde{d}_c$ is evaluated at the charm mass scale% and we have separated $d_n$ into $d_n(d) \pm d_n(w)$, the first contribution comes from the $w$ running into light quarks EDMs and CEDMs, and the the second one from $w$ itself
.
The upper limit $d_n < 2.9 \times 10^{-26}$ then implies
\begin{equation}
\label{eq:CEDMc_bound} 
|\tilde{d}_c| < 1.0 \times 10^{-22} {\rm cm}\,,
\end{equation}
or, equivalently, $m_c |\tilde{d}_c| < 6.7 \times 10^{-9}$. The bound is derived considering only the charm dipole contirubition to $d_n$. %This is to be compared with the previous and only bound existing in the literature, $|\tilde{d}_c| < 3 \times 10^{-14}$ cm, obtained from $\psi^\prime \to \psi \pi^+ \pi^-$ at the Beijing spectrometer \cite{Kuang:2012wp}.
An analysis analogous to the one we performed here can be carried out also for the bottom and top CEDMs, as was done in \cite{Chang:1990jv} and \cite{Kamenik:2011dk}. We give here the bounds one obtains in this way for the heavy quarks, in terms of the coefficients of operators analogous to (\ref{eq:upDipole})
\begin{eqnarray}
\label{EDMu_bounds}
 {\rm Im}(c_{uu})  < 1.3 \times 10^{-8}, &\quad  {\rm Im}(c_{cc})  < 1.8 \times 10^{-5}, \\
 {\rm Im}(c_{dd})  < 8.4 \times 10^{-9}, &\quad   {\rm Im}(c_{bb})  < 1.7 \times 10^{-4}, \\
 &\quad  {\rm Im}(c_{tt})  < 3.3 \times 10^{-2},
 \end{eqnarray}
where all the coefficients are evaluated at the scale $\Lambda = 1$ TeV, and we have shown the $d_n$ bounds on the lighter quarks CEDMs for comparison.

\section{Implications for new physics}
\label{sec:pheno}

As discussed in section \ref{sec:introduction},  the current and foreseen experimental reach in $d_n$ will impact, thanks to the bound we derived, on the flavour violating phenomenology of new physics. More specifically we consider CP violating $D$ meson decays, for which the current measurement of $\Delta$A$_{\rm CP}$ implies ${\rm Im}(c_{uc,cu}) < 3.8 \times 10^{-6} \times (\Lambda /{\rm TeV})^2$. This bound is plagued by $O(1)$ long-distance uncertainties of the matrix elements of the dipole operators.

Property i) of the class of models under considerations makes it convenient to rewrite the coefficients of (\ref{eq:upDipole}) as
\begin{equation}
 c_{ij} = c \,W^L_{i3} W^R_{3j}\,,
 \label{eq:c_Wij}
\end{equation}
where the quantities $W^{L,R}_{k3,3k}$ quantify the communication of the k$^{\rm th}$ generation of quarks with the new physics states associated with the third generation. Many NP models of flavour realise $W^L_{k3} \simeq V^{\rm CKM}_{k3}$, since as we discussed flavour violation in the down quark sector has to be kept under control. With this assumption for the size of $W^L$, the bounds from $d_n$ and $\Delta$A$_{\rm CP}$ imply
 \begin{eqnarray}
\label{eq:EFT_bounds}
\Delta{\rm A}_{\rm CP}:& \quad  |W^R_{3c}| < 1.1 \times 10^{-3}, &\qquad |W^R_{3u}| < 9.2 \times 10^{-5}\,, \\
 d_n:& \quad |W^R_{3c}| < 4.4 \times 10^{-4}, &\qquad |W^R_{3u}| < 3.7 \times 10^{-6}\,.
 \end{eqnarray}
where we have chosen a NP scale $\Lambda = 1$ TeV, and considered one operator at a time.
Without taking into account the contributions from the charm CEDM, one could have saturated the $\Delta$A$_{\rm CP}$ measured value without being in conflict with the EDMs constraints, via requiring a very small $W^R_{3u}$, see e.g. \cite{Giudice:2012qq}. Now this possibility is challenged and, with the foreseen experimental sensitivities, in these models the neutron EDM will become by far the most powerful observable to probe the flavour violating parameters $c_{uc,cu}$.
We stress that all these bounds should be considered as $O(1)$ limits, barring fine-tunings of the unknown coefficients and overall phases in front of the operators, that could potentially affect the above conclusion. %This implies, for example, that formally there is the possibility to make the phases entering the CEDMs small so to be in agreement with the $d_n$ bound, while keeping larger the ones relevant to $\Delta$A$_{\rm CP}$ and invalidate the above conclusion.

\begin{figure}
\begin{center}
  \includegraphics[width=8.9 cm]{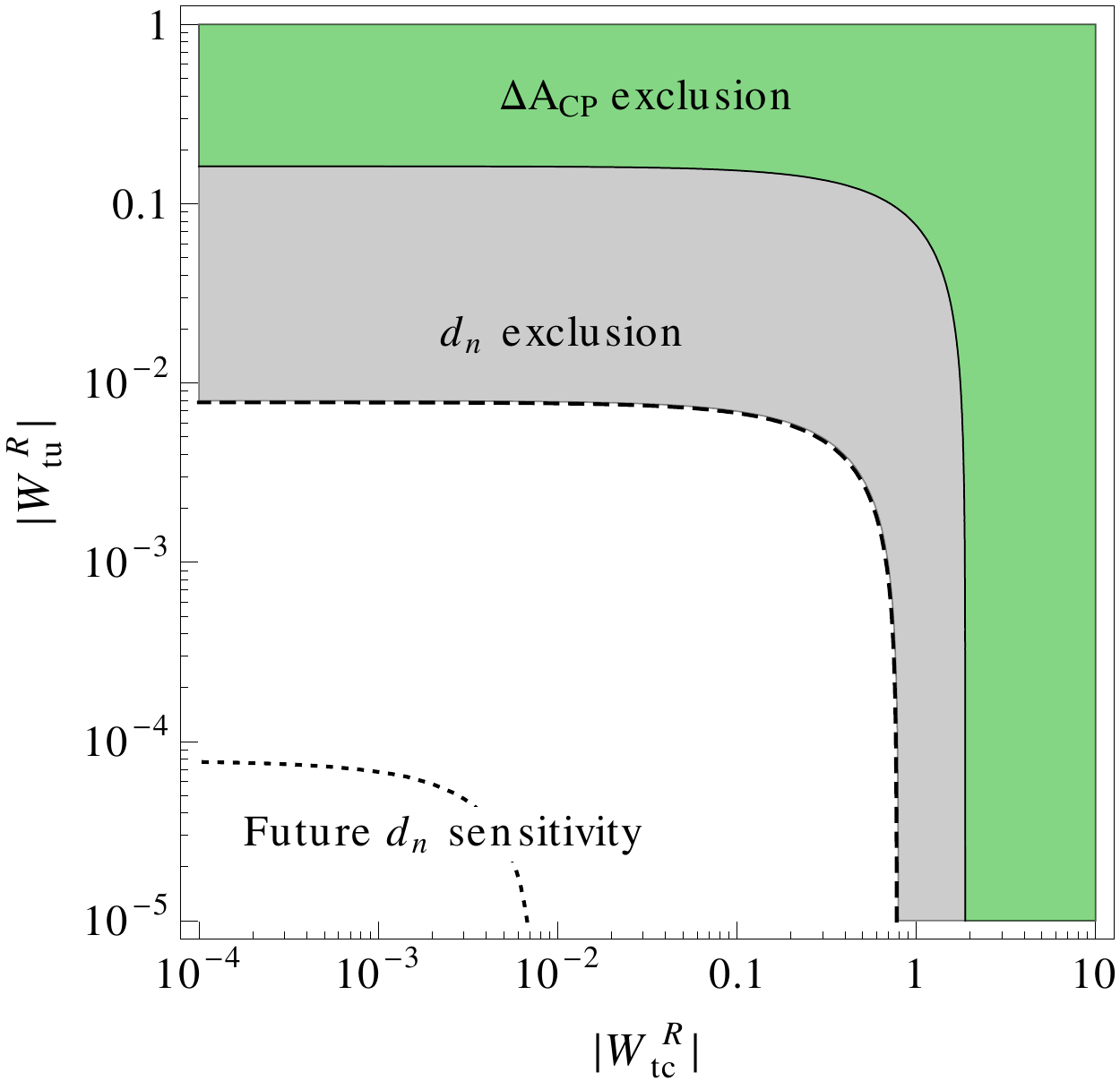}
\end{center}
  \caption{The lines represent the experimental sensitivities to the flavour violating matrix elements in split-families SUSY, the shaded regions are currently excluded. Dashed: current neutron EDM. Dotted: projected neutron EDM. Continuous: Direct CP asymmetry in $D$ decays.}
\label{fig:bounds_SUSY}
 \end{figure}

Supersymmetry with split families \cite{Dimopoulos:1995mi,Cohen:1996vb,Papucci:2011wy} is an explicit realisation of the situation described above. The contributions to the coefficients of eq. (\ref{eq:upDipole}) are dominated by those from gluino-stop loops, which read 
\begin{equation}
\label{eq:Contrib_SUSY}
 \frac{c_{ij}}{\Lambda^2}  = \frac{\alpha_s}{4 \pi} \frac{1}{m_{\tilde g}^2} \frac{A_t - \mu \cot \beta}{m_{\tilde t}} \frac{5}{36}\, g_8(x_{gt})\,W^L_{i3} W^R_{3j},
 \label{SUSY_C8}
\end{equation}
where now $W^{L,R}$ are the mixing matrices entering the gluino-quark-squark vertices of respective chirality, which are responsible for the flavour violation, and $g_8$ is a loop function ($g_8(1)=1$).
It turns out that in split-families SUSY, under some motivated assumptions, one can derive more robust bounds with respect to those analogous to (\ref{eq:EFT_bounds}). In fact one can drop the hypothesis of switching on one operator at a time at the high scale, and instead consider all the leading contributions to $d_n$ at once. To do so it is sufficient to add, to the up and charm CEDMs, only the up quark EDM $d_u$. This does not introduce new parameters, since its form will be analogous to (\ref{eq:Contrib_SUSY}), with the only change of $g_8$. In fact the top and bottom CEDMs contributions are subleading by more than one order of magnitude, once the appropriate Yukawa and mixing suppressions are taken into account.
Moreover it is reasonable to assume that also the down quark EDMs contribution can be neglected: not only they receive a further $y_b/y_t$ suppression with respect to the up and charm ones, but the relevant mixing matrices enter the $\epsilon_K$ parameter, which constrains them to be much smaller than the corresponding ones in the up quark sector, if stops and sbottoms have similar masses.
Note finally that we neglect the contribution to $d_n$ that would come from CP violation in the gaugino and Higgs sectors, which is very strongly constrained by the bound on the electron EDM \cite{Barbieri:2014tja}.
%For a more detailed discussion of th we refer the reader to \cite{Sala:2013osa}.\\
The bounds from $d_n$ and $\Delta$A$_{\rm CP}$ derived in this way are shown in Figure \ref{fig:bounds_SUSY} in the $|W_{tc}^R|$--$|W_{tu}^R|$ plane, for $m_{\tilde{g}} = 2 m_{\tilde{t}} = 1.5$ TeV and $(A_t - \mu/\tan \beta)/m_{\tilde{t}} = 1$.
For illustrative purposes we have assumed the elements of $W^L$ to be equal in magnitude to the respective CKM ones and all the phases to be maximal.%The generalization to the case where there are deviations from these reference values can be inferred from eq. (\ref{eq:Contrib_SUSY}).

Another interesting picture where to study the impact of this bound is the one of composite Higgs models \cite{Kaplan:1983fs,Georgi:1984af,Contino:2003ve,Agashe:2004rs} with partial compositeness \cite{Kaplan:1991dc}, as a concrete realisation of a dynamical suppression of flavour violation.
The careful analysis of \cite{Konig:2014iqa} shows that, thanks to the new bound on $\tilde{d}_c$, it is the neutron EDM that gives the stronger constraints on the masses of the composite partners of the up quarks. For more details on both this case and the SUSY one see \cite{Sala:2013osa}.

\section{Summary and outlook}
\label{sec:outlook}
We have derived a bound on the charm chromo-electric dipole moment $\tilde{d}_c$, via its threshold effect in the three gluon Weinberg operator $w$. This operator in turn contributes to hadronic dipole moments, and the $d_n$ bound yields to $\tilde{d}_c < 1.0 \times 10^{-22}$ cm  at 90\% C.L. at the charm mass scale. %This is one of the two main results of this paper.
If one had neglected the impact of $w$ to $d_n$, the resulting bound would have been $|\tilde{d}_c| < 1.2 \times 10^{-20} {\rm cm}$, coming from the $\tilde{d}_c$ contribution to the light quarks CEDMs. This stresses the importance of reducing the theory uncertainty of the $w$ contribution to hadronic EDMs, whose size is at present debated \cite{Engel:2013lsa}. Another theoretical progress, that is further motivated by this result, is the determination of the long-distance contribution of the charm EDMs to $d_n$.

We also pointed out the relevance of this bound for models allowing for a non-negligible flavour violation in the up quarks sector. These models are still largely unconstrained due to the weakness of the flavour and CP violating bounds compared to those for the down-quark sector. Before this work, the CP asymmetry in flavour violating $D$ decays, $\Delta$A$_{\rm CP}$, was setting the stronger constraints on some relevant flavour violating parameters in these models. We found that the current bound on $d_n$ is already sligthly more constraining than $\Delta$A$_{\rm CP}$. More importantly, the lack of a theoretical understanding of the SM contribution to $\Delta$A$_{\rm CP}$, combined with the expected improvement in experimental sensitivity to $d_n$, will make the neutron EDM the most sensitive probe for these flavour violating parameters, strengthening the current bounds by more than two orders of magnitude.
This conclusion would be improved by a further order of magnitude if the deuteron EDM will be measured with a precision of $\sim 10^{-29} e$~cm, which is the value aimed at by the proposal \cite{bnl:gov}.
This interplay of $d_n$ and flavour violating observables constitutes an additional motivation to achieve a better theoretical control of direct CP violations in $D$ decays.

\section*{Acknowledgments}
I thank the organisers of MoriondEW 2014 for the enjoyable ambience created in La Thuile, as well as for partial support and for the opportunity to give this talk. I thank Michele Papucci for many precious discussions about these subjects. %, as well as Jordy De Vries, Martin Gorbahn, Ulrich Haisch, Martin Jung and Emanuele Mereghetti for useful discussions.
The participation to this conference was also supported by the European Research Council (ERC) under the EU Seventh Framework Programme (FP7/2007-2013) / Erc Starting Grant (agreement n. 278234 - NewDark project).

\section*{References}
\small
\bibliography{Dipoles}
%\begin{thebibliography}{99}
%\bibitem{ja}C Jarlskog in {\em CP Violation}, ed. C Jarlskog
%(World Scientific, Singapore, 1988).

%\bibitem{ma}L. Maiani, \Journal{\PLB}{62}{183}{1976}.

%\bibitem{bu}J.D. Bjorken and I. Dunietz, \Journal{\PRD}{36}{2109}{1987}.

%\bibitem{bd}C.D. Buchanan {\it et al}, \Journal{\PRD}{45}{4088}{1992}.

%\end{thebibliography}

\end{document}